\documentclass[a4paper,11pt]{article}
\usepackage{pos}

\renewcommand{\hookAfterAbstract}{%
  \par\bigskip
  \textsc{P3H-24-059, TTP24-031}
  \href{}
}

\title{Precision Calculations in $B$ physics}

\author*[a]{Matthias Steinhauser}

\affiliation[a]{Institut f{\"u}r Theoretische Teilchenphysik,\\
   Wolfgang-Gaede Stra\ss{}e 1, 76128 Karlsruhe, Germany}

\emailAdd{matthias.steinhauser@kit.edu}

\abstract{ We discuss recent higher order calculations to properties of $B$
  mesons. This includes next-to-next-to-leading order corrections to nonleptonic
  and next-to-next-to-next-to-leading order corrections to semileptonic $B$
  meson decays. The latter is important in connection to the determination of
  the CKM matrix element $V_{cb}$.  We also discuss next-to-next-to-leading
  order corrections to the width difference in the neutral $B$ meson system.
}

\FullConference{42nd International Conference on High Energy Physics (ICHEP2024)\\
18-24 July 2024\\
Prague, Czech Republic\\}

\begin{document}
\maketitle



\section{Introduction}

In this contribution we discuss three physical quantities from $B$
meson physics where higher order corrections are
important: Semileptonic and nonleptonic decays
in Sections~\ref{sec::SL} and~\ref{sec::nonlep}, respectively,
and $B$ meson mixing in Section~\ref{sec::bmix}.


\section{\label{sec::SL}Semileptonic $B$ meson decay}

Semileptonic $B$ meson decays are important ingredients for the determination
of the Ca\-bibbo–Ko\-ba\-ya\-shi–Mas\-kawa matrix elements $V_{ub}$ and
$V_{cb}$. The total decay rate and moments thereof are of particular
importance for the inclusive determination of $|V_{cb}|$. It is thus crucial
to compute higher order corrections to these quantities in order to move
towards a solution of the long-standing tension between the inclusive and
exclusive determinations of $|V_{ub}|$ and $|V_{cb}|$.

If Ref.~\cite{Fael:2020tow} the third-order corrections to the decay rate
$b\to c\ell \nu$ have been computed (see also Ref.~\cite{Czakon:2021ybq} for
partial results). In the following, we briefly outline the main ideas of this
computation.

\begin{figure}[b]
  \begin{center}
  \begin{tabular}{ccccc}
    \raisebox{0.2em}{\includegraphics[width=0.16\textwidth]{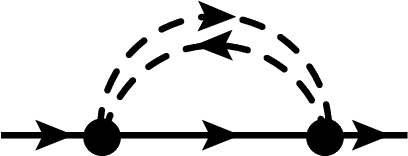}} &
    \includegraphics[width=0.16\textwidth]{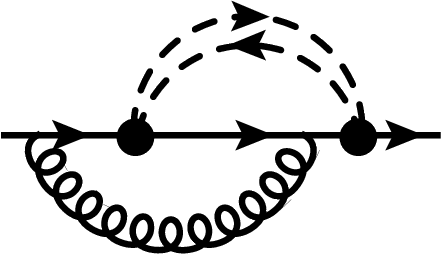} &
    \includegraphics[width=0.16\textwidth]{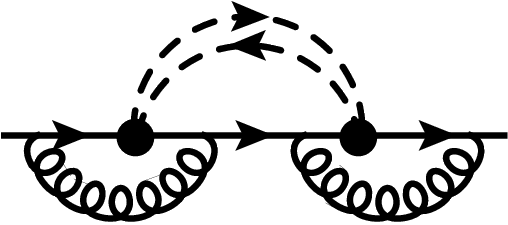} &
    \includegraphics[width=0.16\textwidth]{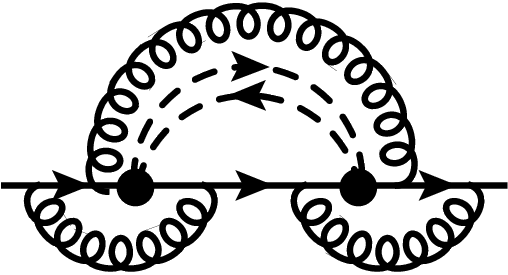} &
    \includegraphics[width=0.17\textwidth]{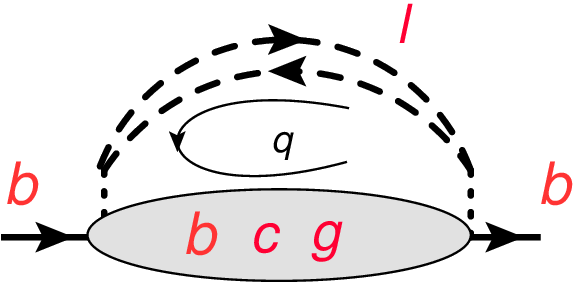}
    \\ (a) & (b) & (c) & (d) & (e)
  \end{tabular}
  \end{center}
  \vspace*{-1em}
  \caption{\label{fig::diags}(a) to (d): Representative Feynman diagrams
  to the semileptonic $B$ meson decay at LO, NLO, NNLO and N$^3$LO.
  (e): generic diagram.}
\end{figure}

A convenient approach to obtain inclusive quantities is based on the optical
theorem. If specified to decay rates one has to compute the imaginary parts of
forward-scattering amplitudes of the decaying particle. In our case we have to
consider corrections up to five-loop order to the bottom quark two-point
function, as can be seen from Fig.~\ref{fig::diags}, where representative
diagrams for the leading (LO), next-to-leading (NLO), next-to-next-to-leading
(NNLO) and next-to-next-to-next-to-leading order (N$^3$LO) corrections are shown.

To date, the techniques to compute five-loop diagrams which depend on two mass
scales ($m_c$ and $m_b$) are not available. The main idea to nevertheless
arrive at precise results is based on expansions in the quantity
$\delta=1-m_c/m_b$, i.e., around the limit $m_c=m_b$. This approach has been
successfully applied at order $\alpha_s^2$ in Ref.~\cite{Dowling:2008mc}.  At
order $\alpha_s^3$ the implementation of this idea consists of the following
steps: The starting point are five-loop diagrams, see Fig.~\ref{fig::diags}.
One performs the loop-integration involving the charged lepton and the
neutrino in $d=4-2\epsilon$ dimensions.  This reduces the problem to four
loops at the price of introducing a massless propagator with momentum $q$
raised to non-integer power (see Fig.~\ref{fig::diags}(e)).  $q$ is an
external momentum of the gray blob which represents the remaining three loop
integrations over $k_i$ ($i=1,2,3$).  At this point, one realizes that all
loop-momenta either scale as $m_b$ (``hard'') or $\delta \cdot m_b$
(``ultra-soft''). Furthermore, $q$ has to be ultra-soft; otherwise there is no
imaginary part.  It is possible to determine the dependence of the gray blob
on $q$ before actually performing the integrations over $k_i$. Thus we can
integrate over $q$ and remain with three-loop integrals, which can be solved
analytically.

\begin{figure}[t]
  \begin{center}
  \begin{tabular}{ccc}
    \includegraphics[width=0.3\textwidth]{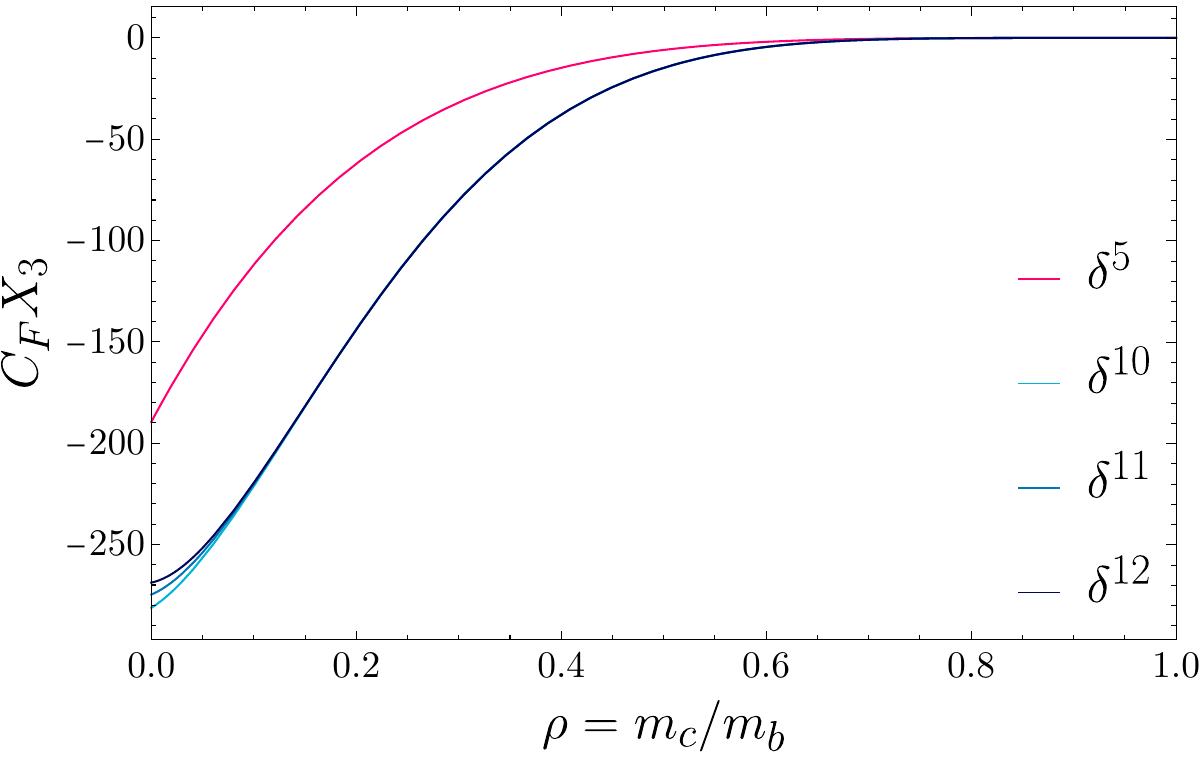} &
    \includegraphics[width=0.3\textwidth]{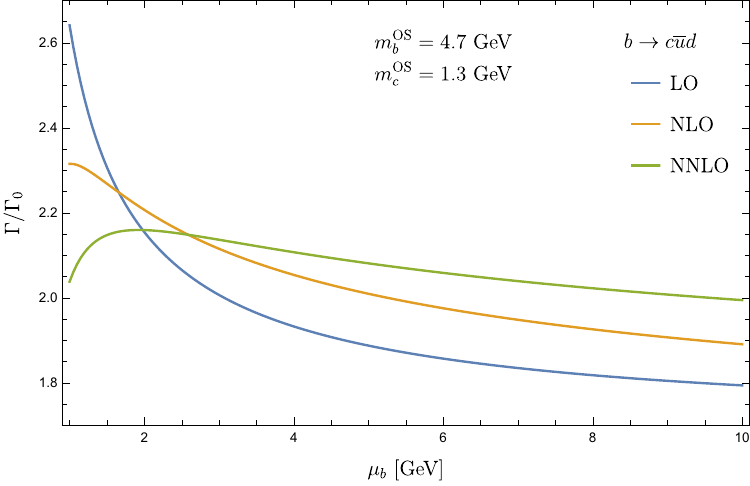} &
    \includegraphics[width=0.3\textwidth]{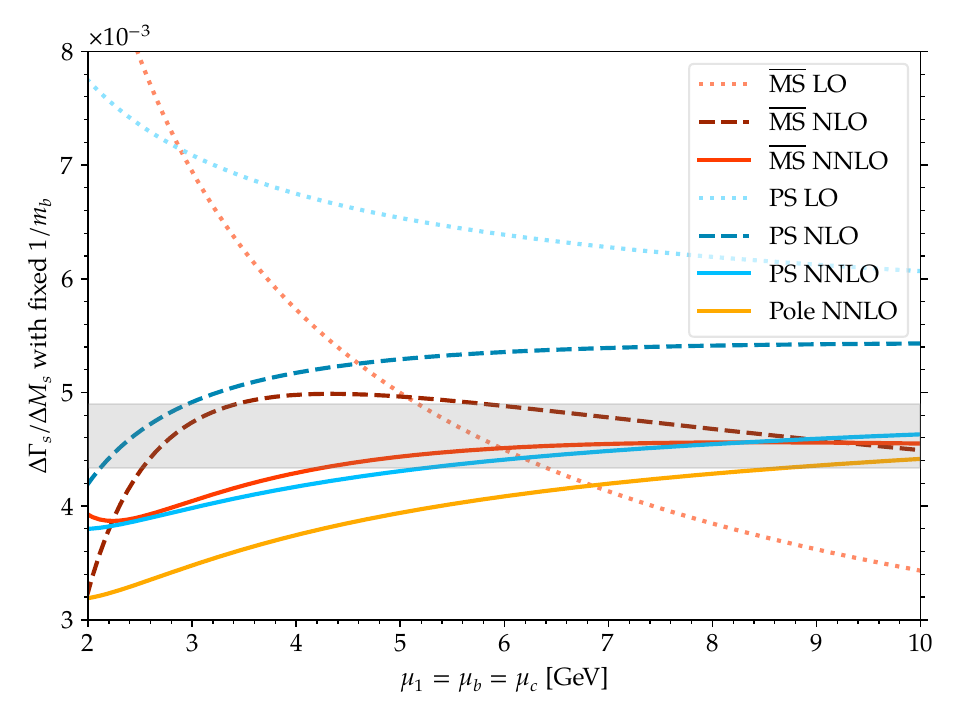}  
  \end{tabular}
  \end{center}
  \vspace*{-2em}
  \caption{\label{fig::X3}Left: Third-order coefficient of the total
    semileptonic decay rate of a $b$ quark as a function of $m_c/m_b$.
    Middle: The dependence of the rate for $b \to c \bar u d$ on the
    renormalization scale $\mu_b$ at LO, NLO and NNLO. Charm and bottom quark
    masses are renormalized in the on-shell scheme.  Right: Renormalization
    scale dependence of $\Delta\Gamma_s/\Delta M_s$ at LO, NLO and NNLO for
    the $\overline{\rm MS}$ and PS scheme. The gray band represents the
    experimental result.}
\end{figure}

In Ref.~\cite{Fael:2020tow} terms up to $\delta^{12}$ have been computed.  The
results are shown in Fig.~\ref{fig::X3} where the coefficient of order
$\alpha_s^3$ is shown as a function of $m_c/m_b$.  One observes a rapid
convergence after including higher order terms in $\delta$ which allows for the
determination of the third-order coefficient with a precision below the percent
level; see Ref.~\cite{Fael:2020tow} for details.

The results of of Ref.~\cite{Fael:2020tow} have been used in recent
determinations of $|V_{cb}|$~\cite{Bordone:2021oof,Bernlochner:2022ucr,Hayashi:2023axn}. In Ref.~\cite{Bordone:2021oof}
it is mentioned that the $\alpha_s^3$ term leads to a reduction
of the uncertainty due to the decay rate by a factor two.
Furthermore, it leads to a shift of $|V_{cb}|$ by only $0.6\%$
which demonstrates that the inclusive determination
of $|V_{cb}|$ is stable w.r.t. higher order
perturbative corrections.

In Ref.~\cite{Fael:2022frj} the same approach as mentioned above has been
applied to obtain results for moments to order
$\alpha_s^3$.  Note, however, the no experimental cuts are applied.
More recently corrections of order $\alpha_s^2$
to the leptonic invariant mass spectrum (``$q^2$ moments'')
have been computed in Ref.~\cite{Fael:2024gyw}.
Recently also the corrections of order $\alpha_s^3$ to $b\to u\ell\nu$
became available:
In Ref.~\cite{Chen:2023dsi} results in the large-$N_c$ limit
have been obtained and in Ref.~\cite{Fael:2023tcv} the fermionic
contributions have been computed.



\section{\label{sec::nonlep}Nonleptonic $B$ meson decay}

Similar to semileptonic $B$ meson decays also the prediction for nonleptonic
decays is organized as a double expansion in $\alpha_s$ and
$\Lambda_{\rm QCD}/m_b$ (see, e.g., Ref.~\cite{Lenz:2022pgw}). Until recently,
the leading term, which describes the partonic decay of the $b$ quark, was
only available to NLO~\cite{Bagan:1994zd,Bagan:1995yf,Krinner:2013cja}.
As a consequence, the by far dominant contribution to the uncertainty of the
theory predictions comes form the variation of the renormalization
scale~\cite{Lenz:2022pgw}.  In Ref.~\cite{Egner:2024azu} the NNLO correction
have been computed\footnote{First NNLO results have been obtained in
  Ref.~\cite{Czarnecki:2005vr}, however, only for massless final-state quarks
  and without the resummation of $\log(m_W/m_b)$ terms.} for various
partonic channels, in particular for $b\to c\bar{u} d$ and $b\to c\bar{c} s$
with one and two massive charm quarks in the final state, respectively.

For the computation the optical theorem has been used, which means that at
NNLO the imaginary parts of four-loop diagrams have to be considered.  There
are contributions with gluon exchanges on the bottom-charm fermion line, which
are in analogy to the semileptonic decay. Furthermore, there are QCD
corrections in the closed quark-antiquark loop and gluon exchanges between the
bottom-charm fermion line and the closed quark-antiquark loop. It is the
latter class of diagrams which makes this calculation more involved than in
the semileptonic decay.  In particular, it is not possible to reduce the
number of integrations in a simple way as in the semileptonic case (see
Section~\ref{sec::SL}) and thus one has to compute four-loop integrals with
two mass scales, $m_c$ and $m_b$. This requires new techniques, which
come with several challenges, see Refs.~\cite{Egner:2023kxw,Egner:2024azu} for
details.

One of the challenges is the computation of the Feynman integrals which depend
on $\rho =m_c/m_b$. 
%
%
%
A further challenge of the NLO calculation is the proper choice of operator
basis such that we are allowed to apply Fierz relations in $d\not=4$
dimensions. This concerns in particular the definition of the evanescent
operators. More details can be found in Ref.~\cite{Egner:2024azu}.

In Fig.~\ref{fig::X3} we show the result for the (normalized) decay rate
as a function of the renormalization scale $\mu_b$. The blue, orange and green
curves represent the LO, NLO and NNLO result. The quark masses are renormalized
on-shell. If we estimate the theory uncertainty from the maximum and minimum
for $\mu_b\in \{m_b/2, 2m_b\}$ (divided by two), we obtain a
reduction from 6.3\% at NLO and to 3.5\% at NNLO.  At the central scale
$\mu_b= m_b$ the ${\cal O}(\alpha_s)$ corrections shift the LO results by
6.5\% and the NNLO induce a further shift of less than 3.5\%.  For a more
phenomenological analysis one has to transform the quark masses
into short-distance schemes which is planned in future work.


\section{\label{sec::bmix}$B$ meson mixing}

The oscillation of a neutral $B_q$ meson (with $q=s,d$) into its anti-particle
is induced by the weak interaction.  Within the Standard Model such
$\Delta B=2$ transitions are mediated by the exchange of $W$ bosons in box
Feynman diagrams.  They contain dispersive and absorptive parts which are
related to the mass and decay matrices ($M^q$ and $\Gamma^q$), which in
turn are related to the experimentally accessible quantities
as, e.g.,
              $\Delta\Gamma_q / \Delta M_q =
              -\mbox{Re} \Gamma_{12}^q / M_{12}^q$.
Here, $\Delta\Gamma_q$ and $\Delta M_q$ are width and mass differences of the
two physical mass eigenstates in the neutral $B$ meson system.
In the following we concentrate on $\Delta\Gamma_s$
where precise experimental results exist
$\Delta\Gamma_s^{\rm exp}  =  (0.082 \pm 0.005)\; \mbox{ps}^{-1}$~\cite{hfag}.


NLO predictions for $\Delta\Gamma_s$ are available since more than 25
years~\cite{Beneke:1998sy,Ciuchini:2003ww,Beneke:2003az,Lenz:2006hd}.
However, only recently NNLO corrections became
available, see~\cite{Gerlach:2022hoj} and references therein.
The theory predictions are based on the construction of
two effective theories: In a first step one integrates out the heavy
degrees of freedom of the Standard Model and arrives at the so-called
$\Delta B=1$ theory. The decay width is obtained from the absorptive part of a
correlator with two $\Delta B=1$ operator insertions.  It is computed by
applying the Heavy Quark Expansion. This leads to effective $\Delta B=2$
operators, which mediate the $B_s-\overline{B}_s$ transition.

At NNLO one has to face several challenges.
Some of them are related to the proper choice of the $\Delta B=2$            
operators, in particular their evanescent contributions.
Furthermore, there are a number of technical challenges.
For example the computation of three-loop Feynman integrals
for finite charm and bottom quark masses,
traces over products with up to 22 $\gamma$ matrices
or tensor integrals up to rank 11.
Their solutions have been discussed at length in a recent
publication~\cite{Reeck:2024iwk}.

In Fig.~\ref{fig::X3} we show the renormalization scale ($\mu_1$)
dependence of the theory prediction of $\Delta\Gamma_s/\Delta M_s$ at LO,
NLO and NNLO for two renormalization schemes, $\overline{\rm MS}$ and PS. One
observes that after the inclusion of higher order corrections the curves become
flatter. Furthermore, the scheme dependence is significantly reduced
and good agreement to the experimental result is observed.
The NNLO result is also shown in the pole scheme, which, however, is not
adequate to describe $\Delta\Gamma_s/\Delta M_s$.


%
%


\section*{Acknowledgements}  

This research was supported by the Deutsche Forschungsgemeinschaft (DFG,
German Research Foundation) under grant 396021762 --- TRR 257 ``Particle
Physics Phenomenology after the Higgs Discovery''.  I would like to thank the
organizers of ICHEP2024 for the posibility to present our results at the
conference. Furhtermore, I would like to thank Manuel Egner, Matteo Fael,
Marvin Gerlach, Ulrich Nierste, Kay Sch\"onwald, Pascal Reeck and Vladyslav
Shtabovenko for the fruitful collaboration on the topics discussed in this
contribution.


\bibliographystyle{jhep} 
\bibliography{inspire_b.bib,inspire_b_manual.bib}

%

\end{document}